\documentclass[nofootinbib,showpacs]{revtex4}

\usepackage{bm}

\begin{document}

\title{Abelian family symmetries and the simplest models \\ that give $\theta_{13}=0$ in the neutrino mixing matrix}

\author{Catherine I Low} \email{c.low@physics.unimelb.edu.au}
\affiliation{School of Physics, Research Centre for High Energy
  Physics, The University of Melbourne, Victoria 3010, Australia}
\date{\today}

\begin{abstract}
I construct predictive models of neutrino mass and mixing that have fewer parameters, both in the lepton sector and overall, than the default seesaw model. The predictions are $\theta_{13}=0$ and one massless neutrino, with the models having a $Z_4$ or $Z_2$ symmetry and just one extra degree of freedom: one real singlet Higgs field.
It has been shown that models with an unbroken family symmetry, and with no Higgs fields other than the Standard Model Higgs doublet produce masses and mixing matrices that have been ruled out by experiment. Therefore, this article investigates the predictions of models with Abelian family symmetries that involve Higgs singlets, doublets and triplets, in the hope that they may produce the maximal and minimal mixing angles seen in the best fit neutrino mixing matrix. I demonstrate that these models can only produce mixing angles that are zero, maximal or unconfined by the symmetry. The maximal mixing angles do not correspond to physical mixing, so an Abelian symmetry can, at best, ensure that $\theta_{13}=0$, while leaving the solar and atmospheric mixing angles as free parameters. 
To generate more features of the best-fit mixing matrix a model with a non-Abelian symmetry and a complicated Higgs sector would have to be used. 
\end{abstract}


\maketitle

\section{Introduction}

The plethora of free parameters is a major reason why the Standard Model (SM) is considered to be only a partial theory. 
Most of the free parameters in the SM are associated with the existence of the three families of fermions. If neutrino mass is included, there are six masses, three mixing angles and one phase in the lepton sector. If neutrinos are Majorana particles there are an additional two phases in the mixing matrix, and if neutrinos gain mass via the seesaw mechanism \cite{seesaw1,seesaw2,seesaw3,seesaw4,seesaw5} with three right handed neutrinos there are another nine free parameters associated with the heavy right handed neutrinos, giving a total of twenty-one parameters associated with the leptons \cite{santamaria,jenkins}. 
If only two right handed neutrinos are present the leptons have fourteen parameters.   

A way to reduce the number of free parameters is to add a family symmetry to the SM. A family symmetry is a particularly intriguing option for the lepton sector because when neutrinos are massless there is an accidental family symmetry $U(1)_{L_e}\times U(1)_{L_\mu}\times U(1)_{L_\tau}$. Since neutrinos are very light, this family symmetry can be considered to be only slightly broken. If there are three massless right handed neutrinos there is also an accidental symmetry $U(3)_{\nu_R}$, which consists of transformations of the right handed neutrinos. It is possible that a subgroup of the $U(1)_{L_e}\times U(1)_{L_\mu}\times U(1)_{L_\tau}\times U(3)_{\nu_R}$ family symmetry remains unbroken once the neutrino mass generation mechanism is added to the minimal Standard Model.

A lepton family symmetry has also been suggested as a way to explain the interesting features of the neutrino mixing matrix.
The mixing matrix, or Maki-Nakagawa-Sakata (MNS) matrix, parameterised by
\begin{equation}
U_\mathrm{MNS}=
\left(\begin{array}{ccc} c_{13} c_{12} & c_{13} s_{12} & s_{13} e^{i \phi} \\ 
-s_{23} s_{13} c_{12}e^{-i \phi}- c_{23} s_{12} & -s_{23} s_{13} s_{12}e^{-i \phi}+ c_{23} c_{12} & c_{13} s_{23} \\
-c_{23} s_{13} c_{12}e^{-i \phi}+ s_{23} s_{12} & -c_{23} s_{13} s_{12}e^{-i \phi}- s_{23} c_{12} & c_{13} c_{23}
\end{array}\right)
\end{equation}
(omitting Majorana phases), has two large mixing angles, and one small. The atmospheric mixing angle, $\theta_{23}$, is maximal at best fit \cite{SKatm2,KAMatm,IMBatm,Soudanatmnew,MACROatm}, and the small mixing angle, $\theta_{13}$, only has an upper bound \cite{CHOOZ}. So, if $\theta_{13}$ is taken to be zero, the best fit mixing matrix has the form
\begin{equation}\label{mnsform}
U_\mathrm{MNS}=\left(\begin{array}{ccc} 
\cos \theta_{12} & \sin \theta_{12} & 0 \\
-\frac{\sin \theta_{12}}{\sqrt{2}}&\frac{\cos \theta_{12}}{\sqrt{2}}&\frac{1}{\sqrt{2}}\\
\frac{\sin \theta_{12}}{\sqrt{2}}&-\frac{\cos \theta_{12}}{\sqrt{2}}&\frac{1}{\sqrt{2}}\end{array}\right),
\end{equation}
where $\theta_{12}$ is the solar mixing angle which is large but not maximal \cite{SNOsolnew,K2Ksol,SKsol2,SAGEsol,HOMESTAKEsol,GALLEXsol,GNOsol}. This mixing matrix pattern may be a result of a family symmetry. If $\sin\theta_{12}=\frac{1}{\sqrt{3}}$, Eq. \ref{mnsform} becomes the tri-bimaximal mixing matrix \cite{hps1,zeehe1}, which hopefully could also arise from a symmetry.

Many authors have created family symmetry extensions to the SM that attempt to explain aspects of the neutrino masses and mixing. Some recent examples are references \cite{kubo,kubohiggspot,grimfav,matbmwithvevs,LavLmodel2}. Common to nearly all the models is an extended Higgs sector, involving additional Higgs doublets, triplets or singlets. (The exception is \cite{LavLmodel2} which has only the SM Higgs and an explicitly broken symmetry). Many of these models require further complexities, including explicit symmetry breaking and relationships between some of the Higgs vacuum expectation values (VEVs), which in some models are justified using an analysis of the Higgs potential.

The aim of my previous work \cite{nogo1,nogo2} was to find out whether the mixing matrix of Eq. \ref{mnsform} can be generated by a simpler model than the family symmetry models already suggested. The approach I took was to investigate the simplest possible family symmetry extensions to the SM, and if these models could not produce the required mixing, more complicated models were investigated.        

It was shown in reference \cite{nogo1} that family symmetry models involving just the SM Higgs field can only give neutrino masses and mixings that are ruled out by experiment, or give mass matrices that are completely unrestricted.
This was done by segmenting the study into non-Abelian and Abelian groups. Non-Abelian groups were easily ruled out, as they forced at least two of the charged leptons to be degenerate. Abelian groups were investigated by looking at all the possible mass matrix textures that can be generated by Abelian groups, and finding their mass eigenvalues and mixings angles. 

Since the simplest family symmetry extensions to the SM cannot provide us with a predictive mixing matrix, any model that does predict the neutrino mixing matrix will require Higgs fields in addition to the SM Higgs doublet. 

The models of \cite{nogo2} involve a number Higgs doublets that transform under the family symmetry. All possible forms of mixing matrices that can be derived from an Abelian symmetry with these Higgs fields were found.  $\theta_{13}=0$ could be produced, but the atmospheric mixing and solar mixing angles could not be constrained by the symmetry. Although these models reduce the number of parameters in the mixing matrix, the total number of free parameters is larger than in models without family symmetries. Also the additional Higgs doublets  may allow flavour-changing neutral currents. Non-Abelian groups were not studied, as they could not be systematically searched easily.

This article continues the search to find a model that can generate the neutrino mixing matrix, by extending the search to involve the possibility of Higgs singlets which can contribute to a Majorana mass term for the right handed neutrinos, and Higgs triplets which can give a left handed neutrino Majorana mass term. Since Dirac neutrinos were fully investigated in reference \cite{nogo2}, this paper assumes neutrinos are Majorana particles. 

This paper is structured as follows. In Sec. \ref{secfamsym} the possible Higgs lepton coupling terms are outlined and the family symmetries are described.
Sec. \ref{abelian} shows that for Abelian symmetries, with any number of Higgs fields, the only best-fit mixing angle that can be produced is $\theta_{13}=0$. 
Sec. \ref{nonab} discusses the possibility of using a non-Abelian family symmetry to constrain the lepton masses and mixing matrix. This section shows that for models with simple Higgs sectors non-Abelian symmetries lead to unrealistic lepton masses. In order to get viable models using non-Abelian symmetries either two Higgs doublets are required, or the neutrinos must gain mass using three mass terms: a Dirac mass term, and Majorana mass terms for both the left and right handed neutrinos.
Sec. \ref{models} searches for the minimal model that produces $\theta_{13}=0$ in the neutrino mixing matrix. Since it has already been shown that models with two Higgs doublets can generate $\theta_{13}=0$ \cite{nogo2}, I search for models that have one Higgs doublet, but may have Higgs triplets or singlets. Sets of Majorana and Dirac mass matrices that produce $\theta_{13}=0$, and are compatible with models that have only one Higgs doublet are listed.
The minimal models that produce $\theta_{13}=0$, and allow the experimentally known values for the other mixing angles, involve the seesaw mechanism and only one real Higgs singlet, in addition to the single SM Higgs doublet. The details of these models are presented in Sec. \ref{favmodels}, and the number of free variables in each of the models is discussed. Sec. \ref{rges} demonstrates that the predictions of these models are unchanged by one-loop renormalisation group running.

\section{The action of the family symmetry}\label{secfamsym}

The neutrino mixing matrix $U_{\mathrm{MNS}}$ arises from the diagonalisation of the neutrino mass matrix $M_\nu$ and the charged lepton mass matrix $M_\ell$: $U_{\mathrm{MNS}}=U_{\ell L}^\dagger U_\nu$, where $U_{\ell L}$ diagonalises $M_\ell M_\ell^\dagger$, and $U_\nu$ diagonalises the Majorana matrix $M_\nu$. The mass matrices arise from the Higgs-lepton couplings, and it is these couplings that are directly affected by the family symmetry. Exactly how the leptons couple to the Higgs fields plays an important role in producing the mixing matrix via a family symmetry.

Charged leptons gain mass through a Dirac Yukawa coupling to doublet Higgs fields: 
\begin{equation}
\overline{L}Y_{\ell\; i} \Phi_i \ell_R,
\end{equation}
where $Y_{\ell\: i}$ are the Yukawa coupling matrices, and $L=(\nu_L, \ell_L)$ is the left handed lepton doublet.
With only Higgs doublets, neutrinos can gain Dirac masses via the term
\begin{equation}
\overline{L} Y_{\nu\; i} \Phi^c_i \nu_R,
\end{equation}
Majorana masses via a dimension-5 operator \cite{dim5} from
\begin{equation}
\frac{1}{\Lambda}\overline{L} \Phi^c_i \kappa_{ij} \Phi^{c T}_j L^c,
\end{equation}
and Majorana masses via the seesaw mechanism \cite{seesaw1,seesaw2,seesaw3,seesaw4,seesaw5} when the right handed Majorana neutrino mass comes from a bare mass term:
\begin{equation}
\overline{(\nu_R)^c} M_R \nu_R.
\end{equation}
 Possible mixing matrices arising from a family symmetry and these three neutrino mass terms have been investigated in \cite{nogo1} and \cite{nogo2}. 

Other ways of getting neutrino masses include the left handed leptons coupling to triplet Higgs fields $\Delta_i$ \cite{triplet}:
\begin{equation}
\overline{L}\lambda_i \Delta_i L^c, 
\end{equation}
where $\lambda_i$ are $3\times 3$ coupling matrices.
If the neutral components of the triplets gain VEVs, they will generate a neutrino mass matrix given by $M_\nu=\lambda_i \langle \Delta^0_i \rangle$.
The seesaw mechanism can be altered by adding Higgs singlets $\chi_i$ to give a Majorana mass matrix to the right handed neutrino via 
\begin{equation}
\overline{(\nu_R)^c}\epsilon_i \chi_i \nu_R,
\end{equation}
 so the right handed Majorana matrix is $M_R=\epsilon_i \langle \chi_i \rangle +M_{Bare}$. If the singlets are complex, the neutrinos will also couple to the complex conjugate of the singlets $\chi^*_i$.
Any combination of these mass generation mechanisms can be used to give neutrinos mass. The most general light neutrino mass matrix is given by $M_{\nu} = M_L - M_{Dirac} M_R^{-1} M_{Dirac}^T$, where $M_L$ is the mass term from the triplet Higgs and/or the dimension-5 operator, $ M_R$ is from the bare mass term and/or the singlet Higgs coupling term, and $M_{Dirac}$ is the Dirac mass matrix due to Higgs doublets.

The family symmetry alters the mass terms by ensuring each mass term is invariant under the transformations which act on the different families of Higgs fields and leptons. 
The leptons transform via
\begin{equation}
L \rightarrow X_L L, \: \ell_R\rightarrow X_{\ell R} \ell_R,\: \nu_R\rightarrow X_{\nu R} \nu_R,
\end{equation} 
where the $X$ matrices are sets of $3\times 3$ unitary transformation matrices in family space. 
The families of Higgs doublets, triplets and singlets transform via 
\begin{equation}
\Phi_i \rightarrow V_{\Phi ij} \Phi_j,\:
\Delta_i \rightarrow V_{\Delta ij} \Delta,\:
\chi_i \rightarrow V_{\chi ij} \chi_j,
\end{equation}
where the $V$ matrices are unitary transformation matrices in Higgs family space.
The transformation matrices for each type of field form a representation of the symmetry group. 

These symmetries dictate the form of the Higgs-lepton coupling matrices, and therefore dictate the form of the mass matrices, and hence the mixing matrix pattern. The remainder of this article investigates ways that combinations of mass terms and symmetries can generate allowed forms of the neutrino mixing matrix.

\section{Abelian family symmetries}\label{abelian}

To exhaustively search through all the possible family symmetries all groups need to be examined, and all $3\times 3$ matrix representations of these groups need to be considered as candidates for the transformation matrices. However there are simplifications that make the systematic study of Abelian groups possible. This section demonstrates that all possible mixing matrix forms that can be created by an Abelian symmetry can be generated from mass matrices containing texture zeros, and shows that the only aspect of the best fit neutrino mixing matrix (Eq. \ref{mnsform}) that can be created by an Abelian symmetry is $\theta_{13}=0$.

For each group there are many equivalent representations, but fortunately only a finite number of non-equivalent representations. Two matrix representations $X_i$ and $Y_i$, are considered to be equivalent if they are related by a similarity transformation $X_i=S^\dagger Y_i S$, where $S$ is a unitary matrix. Appendix \ref{app} demonstrates that two different, but equivalent, choices of representation for the lepton transformations yield exactly the same mixing matrix predictions. Choosing a different equivalent representation is identical to choosing a different weak basis.
Every Abelian representation is equivalent to a representation where all the transformation matrices are diagonal, so to study Abelian groups only diagonal $X_L$, $X_{\ell R}$ and  $X_{\nu R}$ matrices need to be considered. These diagonal lepton transformations either dictate that an element of a mass matrix is zero, or the element is completely unrestricted by the symmetry. So the study of Abelian groups can be reduced to the study of mass matrices with texture zeros where all the non-zero elements are unknown. In fact Grimus et al. have shown that all possible textures of this type can be created from a diagonal Abelian symmetry if there are enough Higgs fields in the model \cite{grimlavtextures}.

To demonstrate that diagonal representations give texture zeros consider the $ij$th element of a neutrino mass matrix arising from a Higgs triplet. 
\begin{equation}
M^{ij}_{\nu}=\lambda^{ij}_\alpha \langle \Delta_\alpha \rangle,
\end{equation}
 where $\lambda^{ij}_\alpha$ is the $ij$th element of the Higgs-lepton coupling matrix that couples the $\alpha$th Higgs triplet to the leptons. 
A symmetry will ensure 
\begin{equation}
\lambda^{ij}_\alpha=\sum_\beta X_L^{\dagger ii}V_{\Delta \alpha \beta} \lambda^{ij}_\beta X^{* jj}_L, 
\end{equation}
(no summation over $i$ and $j$) showing that the matrix element $\lambda^{ij}$ for a particular Higgs field can only be related to $\lambda^{ij}$ for other Higgs fields, and not to other elements of $\lambda$. Therefore, the $ij$th element of the mass matrix $M_{\nu}$ cannot be related to any other element (aside from the symmetric relation $M_{\nu ij}=M_{\nu ji}$). So apart from the fact that Majorana mass matrices are symmetric, all non-zero elements of $M_{\nu}$ are completely unrestricted by the symmetry.   
Similarly, Dirac mass matrices for the charged leptons and the neutrinos will get texture zeros from the Abelian symmetry, as will the right handed neutrino mass matrix, and neutrino mass matrices due to a dimension-5 operator. 

Note that when the neutrino mass matrix arises from the seesaw mechanism, it is the constituent matrices ($M_L$, $M_{Dirac}$ and $M_R$) that gain texture zeros from the symmetry. The resultant left handed neutrino mass matrix given by $M_\nu=M_L-M_{Dirac} M_R^{-1}M_{Dirac}^T$, may not have texture zeros, and the non-zero elements may be related to each other. Most systematic studies of neutrino textures consider the textures zeros in $M_\nu$ only, for example \cite{xingtext1,xingtext2,bandotext1,bandotext2}(or in the case of \cite{lavtext} the textures of $M_\nu^{-1}$), so Abelian symmetries can recover mass matrix patterns in addition to those already studied in these articles.

To find the form of the mixing matrices that can arise from Abelian symmetries a computerised search was undertaken that investigated the mixing matrix predictions of all possible mass matrix textures. A similar search was done in \cite{nogo2}, but that search did not allow for textures arising from transforming Higgs triplets and singlets, so it searched through a smaller number of mass matrix patterns.

For a given texture of the neutrino mass matrices and charged lepton mass matrix, sets of matrices were created by inserting random numbers in the matrix whenever a non-zero element appeared, and both the neutrino and charged lepton mass matrices were diagonalised to find the diagonalisation matrices $U_\nu$ and $U_{\ell L}$. 
The neutrino mixing matrix was then created: $U_{\mathrm{MNS}}=U^\dagger_{\ell L} U_\nu$. 

This was repeated using the same textures, but different random numbers for the non-zero elements. Then the two mixing matrices arising from the same texture were compared. If the two mixing matrices had any identical mixing angles then the mixing angle was a result of the symmetry. Many textures can create two or three zero mixing angles, so I specifically searched for mixing matrices with $\theta_{13}=0$ and two non-zero mixing angles, and also mixing matrices that have mixing angle values that are dictated by the symmetry and are non-zero.
This search was performed using textures arising from both seesaw and non-seesaw Majorana mass matrices. For the seesaw cases, three right handed neutrinos were in general assumed,
however all $M_\nu$ matrices arising from fewer right handed neutrinos were automatically created. This is because models with two right handed neutrinos give identical $M_\nu$ matrices to models with three right handed neutrinos when the third column of $M_{Dirac}$ has all zero elements, and $M_R$ is either of an upper $2 \times 2$ block form, or a diagonal form. Models with one right handed neutrino give identical $M_\nu$ matrices as the three neutrino models with two zero columns in $M_{Dirac}$ and a diagonal $M_R$.

The results of this search showed that there are four possible types of mixing angle predictions from an Abelian symmetry. The mixing angles can be zero, maximal, related to masses, or unrestricted by the symmetry. However, the maximal mixing is always unphysical as the two maximally mixed neutrinos are degenerate, as explained in the following paragraph. In models where the mixing angles can be related to the masses, the masses are unrestricted by the symmetry, therefore it is not possible for an Abelian symmetry to fully predict any physically meaningful angle except for a zero angle. This means that values for the solar and atmospheric mixing angles cannot be derived from an Abelian symmetry, though the symmetry may dictate that the angles are related to the mass eigenvalues. 

The maximal mixing angle arises when the Majorana neutrino mass matrix $M_\nu$ has one of two forms. 
\begin{equation}
M_\nu=\left(\begin{array}{ccc}X&0&0\\0&0&X\\0&X&0\end{array}\right),
\end{equation}
where the $X$ elements correspond to unknown parameters, is diagonalised by
\begin{equation}\label{maxatm}
U_\nu=\left(\begin{array}{ccc}1&0&0\\0&\frac{1}{\sqrt{2}}&-\frac{1}{\sqrt{2}}\\0&\frac{1}{\sqrt{2}}&\frac{1}{\sqrt{2}}\end{array}\right), 
\end{equation}
which indicates the second and third neutrino are maximally mixed. Row and column interchanges of this mass matrix also give diagonalisation matrices with one maximal mixing angle. 
This mass matrix describes one Majorana and one Dirac neutrino, or equivalently, one unmixed Majorana neutrino and two maximally mixed Majorana neutrinos with $m_1=-m_2$. Since $\delta m^2=0$ the maximal mixing does not give oscillations. In order to get physical mixing from this mass matrix symmetry breaking terms could be added to break the degeneracy, which will also deviate the mixing from maximal. 

The mass matrix 
\begin{equation}
M_\nu=\left(\begin{array}{ccc}0&X&X\\X&0&0\\X&0&0\end{array}\right)
\end{equation}
has similar properties. It is diagonalised by 
\begin{equation}\label{maxsol}
 U_\nu=\left(\begin{array}{ccc} -\frac{1}{\sqrt{2}}e^{-i \sigma} & \frac{1}{\sqrt{2}}e^{-i \sigma} & 0 \\ 
\frac{\sin \theta}{\sqrt{2}} e^{-i \sigma} & \frac{\sin \theta}{\sqrt{2}} e^{-i \sigma} & \cos \theta \\ 
\frac{\cos \theta}{\sqrt{2}} & \frac{\cos \theta}{\sqrt{2}} & -\sin \theta e^{i \sigma}
\end{array} \right).
\end{equation}
Row and column interchanges of this mass matrix also give a maximal mixing angle. Again the maximally mixed neutrinos $\nu_2$ and $\nu_3$ have $\delta m^2=0$, so maximal oscillation is not possible.

Even if the degeneracy is broken, the maximally mixed neutrinos 
cannot correspond to maximal atmospheric mixing as no combination of possible $U_\ell$'s and the above $U_\nu$'s will have the maximal mixing appearing in the correct place in the mixing matrix. This is true whether or not $\theta_{13}=0$ (see \cite{nogo2} for a more detailed explanation).

Since neither the maximal atmospheric mixing angle nor a particular allowed solar mixing angle can be generated by mass matrices with texture zeros, the only aspect of the best-fit neutrino mixing matrix that can be produced by an Abelian symmetry is $\theta_{13}=0$. For a review of other ways to get a small $\theta_{13}$ see reference \cite{joshtheta13}.

\section{Non-Abelian family symmetries}\label{nonab}

Since Abelian symmetries cannot generate maximal atmospheric mixing or specify an allowed value for the solar mixing, it is worth studying non-Abelian symmetries.
This article does not demonstrate which kinds of mixing angles can be produced by a non-Abelian symmetry, as I did not find a systematic way to investigate all non-Abelian symmetries. However, there are a number of classes of models that will always predict unrealistic masses, and therefore cannot produce viable masses and mixings. These classes are described in this section. Future model building searches should focus on the classes of models that are not ruled out by their mass spectra.

If all the leptons transform via an Abelian symmetry, but the Higgs fields transform via a non-Abelian symmetry, the model will produce mass matrices that have already been considered in Sec. \ref{abelian}. This is because the lepton transformations can always be chosen to be diagonal, and these transformations can only act to force elements of the mass matrices to be zero or unrestricted, regardless of how the Higgs fields transform. Therefore these models offer no advantages over Abelian groups in predicting the neutrino mixing matrix.

Reference \cite{nogo1} considered non-Abelian symmetries with models that have only one Higgs doublet. It was shown that if the charged leptons transform under a non-Abelian family symmetry, the symmetry ensures that at least two charged leptons are degenerate, in stark contrast to the known masses which have a large hierarchy.  So models with only one Higgs doublet are restricted to having the left handed leptons and right handed charged leptons transform under an Abelian group. Or equivalently, the sets of transformation matrices $\{X_L\}$ and $\{X_{\ell R}\}$, must commute. 
If there is more than one Higgs doublet the symmetry may not result in degenerate masses, and may provide a viable model.

This also leaves open the possibility of models with one Higgs doublet that have the right handed neutrinos transforming under an non-Abelian symmetry, while the left handed leptons and right handed charged leptons have commuting transformations.  In this case, if the neutrinos are Dirac particles, the restriction on $M_\nu M_\nu^\dagger$ from the left handed transformation dictates that the mixing matrix is not allowed by experiment \cite{nogo1}, and the right handed neutrino transformation will only further restrict the mixing matrix. Models with seesaw neutrinos and no transforming Higgs are also ruled out by the left handed transformation alone \cite{nogo1}.

Seesaw models with one Higgs doublet and transforming Higgs singlets can also give unrealistic results. If we assume that there are no triplet Higgs fields and Dimension-5 mass terms are not allowed, then the neutrino mass matrix is given by $M_\nu=-M_{Dirac} M_R^{-1} M_{Dirac}^T$. The constraints on the Dirac mass matrix are enough to show that three flavour oscillation is not possible. Appendix \ref{appnonAb} demonstrates that in models where there is only one Higgs doublet, an Abelian left handed transformation and a non-Abelian right handed transformation leads to $M_{Dirac}$ having at least two zero eigenvalues. The resultant neutrino mass matrix $M_\nu$ will also have at least two zero eigenvalues, corresponding to two massless neutrinos, and as $\delta m^2$ will be forced to be zero, any mixing between the degenerate neutrinos is unphysical.

However, if the neutrinos gain mass via the general seesaw mechanism $M_\nu=M_L-M_{Dirac} M_R^{-1} M_{Dirac}^T$, the $M_L$ component can provide additional eigenvalues for $M_\nu$, and may produce acceptable masses and mixing angles.

Therefore, in order to use a non-Abelian group to generate lepton masses and mixing one of two options is required:
\begin{enumerate}
\item A model with two or more transforming Higgs doublets. 
\item A model where the right handed neutrinos transform according to a non-Abelian group, but all the other leptons transform according to an Abelian group, and the neutrinos gain mass via
 a general seesaw mechanism involving Dirac mass terms, Majorana mass terms for both the left and right handed neutrinos and transforming Higgs singlets and/or triplets.
\end{enumerate}
Both of these options require a significant increase in the complexity of the Higgs potential, so it is likely that the models may end up having more parameters in total than the SM.

\section{The search for simple models with $\theta_{13}=0$}\label{models}
 
Section \ref{abelian} showed that Abelian family symmetries can, at best, only ensure that $\theta_{13}=0$, so this section investigates the simplest models that make this prediction. Reference \cite{nogo2} demonstrated that $\theta_{13}$ can be produced by models with two Higgs doublets and a $Z_4$ symmetry. However, this additional Higgs doublet can cause unwanted flavour changing neutral currents, and due to the more complex Higgs potential the models have more free variables than the SM with neutrino masses and no family symmetry. 
To search for simpler models than the two Higgs doublet cases, I looked for models that involve only one Higgs doublet, but one or more Higgs singlets or Higgs triplets. Not all textures can be created by such a model, in particular the Dirac mass matrices must obey the symmetry restrictions
$M_\ell=X_L^\dagger M_\ell X_{\ell R}$ and $M_{Dirac}=X_L^\dagger M_{Dirac} X_{\nu R}$.    

There are three ways combinations of diagonalisation matrices with zero or unrestricted angles can give $\theta_{13}=0$ while leaving the other mixing angles non-zero.
\begin{enumerate}
\item $U_{\nu}$  has an upper left $2\times 2$ block, and $U_{\ell L}$ has a lower right $2\times 2$ block.

\item $U_{\nu}$  is either diagonal, or has an upper left $2\times 2$ block, and $U_{\ell L 31}=0$.

\item $U_{\nu 13}=0$, and $U_{\ell L}$ is either diagonal or has a lower right $2\times 2$ block.

\end{enumerate}

Note that a simultaneous interchange of the rows of $U_{\ell L}$ and the rows or $U_\nu$ corresponds to an interchange of the weak basis leptons, and does not alter the mixing matrix, as $U_{\ell L}^\dagger U_\nu$ remains the same. So stating that $U_{\nu}$ has a lower left $2\times 2$ block and $U_{\ell L}$ has an upper right $2\times 2$ block is equivalent to stating option one above. 

Using the results of the computerised search, this section finds the mass matrices that can generate a mixing matrix with $\theta_{13}=0$ using these three diagonalisation matrix patterns. Symmetry considerations demonstrate that the first two options cannot be created using models with only one Higgs doublet, but the third option, with $U_{\nu}$ having one zero, can be generated.

\subsection{Two $2\times 2$ diagonalisation matrices}

This diagonalisation pattern requires $M_\nu$ to have an upper left $2\times 2$ block form, and $M_\ell M_\ell^\dagger$ to have a lower right $2 \times 2$ block form. To achieve this $M_\ell$ must have at least one column where both the 2nd row and the 3rd row have non-zero elements. 
To get this from a diagonal transformation restriction of $M_\ell = X_L^\dagger M_\ell X_{\ell R}$, $X_L$ must have the form
\begin{equation}
X_L=diag(e^{i\alpha_1}, e^{i\alpha_2}, e^{i\alpha_2}),
\end{equation}
ensuring that rows 2 and 3 have identical patterns of zero and unrestricted elements. This $X_L$ form also puts restrictions on the neutrino mass matrix.

Looking at the non-seesaw neutrino mass matrices first, the restriction on the neutrino-Higgs triplet coupling matrices is $\kappa_i = V_{\Delta ij} X_L^\dagger \kappa_j X_L^*$, where the triplets transform as $\Delta_i \rightarrow V_{\Delta ij}  \Delta_j$. 
If 
\begin{equation}
\kappa_j= \left(\begin{array}{ccc}A & B & C \\ B & D & E \\ C& E & F\end{array}\right)
\end{equation}
then
\begin{equation}
X_L^\dagger \kappa_j X_L^* = \left(\begin{array}{ccc}e^{-2 i \alpha_1 }A &e^{-i( \alpha_1+\alpha_2) } B & e^{-i (\alpha_1+\alpha_2) }C \\e^{-i (\alpha_1+\alpha_2) } B & e^{-2 i\alpha_2 }D &  e^{-2 i\alpha_2 }E \\ e^{-i (\alpha_1+\alpha_2) }C&  e^{-2 i\alpha_2 }E & e^{-2 i\alpha_2 } F \end{array}\right).
\end{equation}
Note that the phases on $B$ and $C$ are the same, as are the phases on $D$, $E$, and $F$. Therefore, the symmetry either forces both $B$ and $C$ to be zero, or allows both $B$ and $C$ to be unconstrained. $D$, $E$, and $F$ are similarly linked. So $M_\nu$ will be made up of some linear combination of 
\begin{equation}\label{2by2blockkappa}
\left(\begin{array}{ccc}X & 0 &0 \\ 0&0&0\\0&0&0\end{array}\right),\:
\left(\begin{array}{ccc}0&X&X\\X&0&0\\X&0&0\end{array}\right)\textrm{ or }
\left(\begin{array}{ccc}0&0&0\\0&X&X\\0&X&X\end{array}\right),
\end{equation}
and could never be of an upper left $2\times 2$ block form. Therefore models with one Higgs doublet and a number of Higgs triplets cannot generate an upper left $2\times 2$ block $U_{\nu}$ and at the same time a lower right $2\times 2$ block for $U_{\ell L}$.

For models with the seesaw mechanism a similar problem occurs. Rows 2 and 3 of $M_{Dirac}$ must have identical patterns due to the form of $X_L$, 
 giving a resultant left handed mass matrix from $M_\nu=-M_{Dirac} M_R^{-1} M_{Dirac}^T$ which also has the same pattern after row 2 and 3 and column 2 and 3 interchanges. So the mass matrix also must have the form of some linear combination of the matrices in Eq. \ref{2by2blockkappa} and cannot produce an upper left $2\times 2$ block matrix. 

So there is no way that an Abelian symmetry with one Higgs doublet can give us $\theta_{13}=0$ via two $2\times 2$ diagonalisation matrices.
However if there is more than one Higgs doublet in the model, then an Abelian symmetry can generate this pattern \cite{nogo2},\cite{matheta13}. 

\subsection{$U_{\ell L}$ with one zero element} 

The program outlined in Sec. \ref{abelian} generated all Dirac mass matrices made of zero and unrestricted elements, diagonalised these matrices, and all the diagonalisation matrices with one zero element were selected. The only Dirac mass matrices that can produce such a diagonalisation matrix are
\begin{equation}\label{onezeroml}
\left(\begin{array}{ccc}0&0&X\\0&0&X\\0&X&X\end{array}\right),\qquad \left(\begin{array}{ccc}0&0&X\\0&0&X\\X&X&X\end{array}\right),
\end{equation}
and row and column interchanges of these matrices. Neither of these matrices can be created from a diagonal family symmetry restriction with one Higgs doublet. To illustrate this consider the diagonal transformations $X_L=diag(e^{i\alpha_1},e^{i \alpha_2}, e^{i\alpha_3})$, $X_{\ell R}=diag(e^{i\beta_1},e^{i \beta_2}, e^{i\beta_3})$.  The constraint on the charged lepton mass matrix $M_\ell = X_L^\dagger M_\ell X_{\ell R}$ reduces to $M^{ij}_\ell =  e^{i (\beta_j-\alpha_i)} M^{ij}_\ell$ . If $M_{\ell 23}$, $M_{\ell 32}$ and $M_{\ell 33}$ are non-zero (as is the case for the matrices in Eq. \ref{onezeroml}) this means the phases $e^{i (\beta_2-\alpha_3)}=e^{i (\beta_3-\alpha_2)}=e^{i (\beta_3-\alpha_3)}=1$, and $\beta_3=\alpha_3=\beta_2=\alpha_2$. Consequently, the phase $e^{i (\beta_2-\alpha_2)}$ will also equal $1$, and there will be no restrictions on the element $M_{\ell 22}$ either, indicating that this element is unknown, and is not constrained to be zero by the symmetry.
Therefore, the above mass matrices that give $U_{\ell L}$ with one zero element cannot arise from a model with one Higgs doublet.

\subsection{$U_\nu$ with one zero element from a triplet Higgs coupling}

The last possibility for finding a mixing matrix with $\theta_{13}=0$ is from models that have $U_\nu$ with one zero element, and $U_{\ell L}$ either a diagonal or of a lower right $2\times 2$ block form. I separate the study of these models into those that can be generated by neutrinos with couplings to triplet Higgs bosons, and those that arise from the seesaw mechanism.

When neutrinos gain mass from a triplet Higgs coupling, the neutrino mass matrix will be a symmetric matrix with elements either zero or unrestricted.
The computer-aided search demonstrated that only two of these mass matrices textures give a diagonalisation matrix with $U_{\nu 13}=0$: 
\begin{eqnarray}
M_\nu&=&\left(\begin{array}{ccc}0&X&X\\X&0&0\\X&0&0\end{array}\right),\label{maj1}\\
M_\nu&=&\left(\begin{array}{ccc}X&X&X\\X&0&0\\X&0&0\end{array}\right).\label{maj2}
\end{eqnarray}
Both of these matrices can be created by symmetries that also generate $M_\ell M_\ell^\dagger$ in a lower right $2\times 2$ block, so $U_{\nu 13}=0$ will result in $\theta_{13}=0$ when the mixing matrix is constructed.
The matrix in Eq. \ref{maj1} can be generated by models with just one SM Higgs doublet through a dimension-5 mass term or by using the seesaw mechanism with a bare mass term. Eq. \ref{maj1} can also be created using one non-transforming Higgs triplet \cite{nogo1}. This matrix can be generated by the $U(1)$ symmetry $L_e-L_\mu-L_\tau$, but if the symmetry is not broken the matrix predicts maximal solar mixing, but no oscillation since $\delta m^2_{sol}=0$, and also predicts $\nu_3$ to be massless. This symmetry has been considered by many authors as a broken symmetry (for examples see \cite{rojL,lavL,grimlavL,barbieriL,petcovL,leungL,joshipuraL}). However, this symmetry acts only on the leptons, and not on any Higgs fields, so the SM Higgs VEV will not spontaneously break the symmetry. The symmetry has to be either broken explicitly, or part of a larger extension to the SM involving other fields to do the symmetry breaking.

The matrix in Eq. \ref{maj2} 
also gives $U_{\nu 13}=0$, and predicts a massless third neutrino. In this mass matrix $\delta m^2_{12}$ and the solar mixing angle are related in such a way that the experimental values for these parameters cannot both be satisfied. One of the three free variables in the mass matrix supplies a contribution to the atmospheric mixing angle, and the other two free variables dictate two masses and the solar mixing angle. The third neutrino is massless. If the mass matrix elements are chosen to satisfy the measured $\delta m^2$ values, the solar mixing will be very close to maximal, and not allowed by experiment. See reference \cite{nogo2} for further explanation.  

This matrix can be created with two Higgs triplets using the symmetry $U(1)_{L_e}\times U(1)_{L_\mu+L_\tau}$, where one of the triplets has a leptonic charge of $L_e=2$ and $L_\mu+L_\tau=0$ and the other a charge of  $L_e=1$, $L_\mu+L_\tau=1$. 

Hence, family symmetry models with triplet Higgs fields can generate $\theta_{13}=0$ via neutrino diagonalisation matrices with one zero element. However, these models have solar mixing angles and mass-squared-differences that are not allowed by experiment.

\subsection{$U_\nu$ with one zero from a seesaw mechanism and singlet Higgs couplings}\label{onezeronu}

The seesaw mechanism generates more possibilities for the neutrino mixing matrix form than triplet Higgs couplings. As well as neutrino mass matrices that contain zero and unrestricted elements, seesaw mass matrices can have related non-zero elements. So the seesaw mechanism with singlet Higgs fields can create mass matrices with no zero elements that are diagonalised by matrices with one zero element. Unlike the mass matrices for Eqs. \ref{maj1} and \ref{maj2}, these matrices do not overconstrain the solar oscillation. 

There are a large number of combinations of Dirac and right handed Majorana neutrino mass matrices that can produce a $U_\nu$ with one zero element. Instead of listing them all, I use some symmetry considerations to omit sets of matrices that cannot be produced by a symmetry in models with only one Higgs doublet.

First of all, if the one zero element in the diagonalisation matrix is to correspond to $U_{e3}$, $M_\ell M_\ell^\dagger$ has to be either diagonal, or of a lower right 
$2\times 2$ block. For $X_L=diag(e^{i\alpha_1},e^{i\alpha_2},e^{i\alpha_3})$, this means that there must be some $X_L$ where $\alpha_1\neq\alpha_2$, otherwise $M_\ell M_\ell^\dagger$ would not satifsy this requirement. This restriction eliminates 
 all neutrino Dirac mass matrices with the same pattern in rows 1 and 2. 

There are two remaining Dirac mass matrices, that combined with $3\times 3$ right handed Majorana mass matrices can give $\theta_{13}=0$:
\begin{eqnarray}\label{favdirac}
M_{Dirac\;1}&=&\left(\begin{array}{ccc}X&0&0\\0&X&0\\0&X&0\end{array}\right),\\
M_{Dirac\;2}&=&\left(\begin{array}{ccc}X&0&X\\0&X&0\\0&X&0\end{array}\right).\label{favdirac2}
\end{eqnarray} 
Note that $M_{Dirac}$ of Eq. \ref{favdirac} has a column of zeros. This means the third right handed neutrino does not couple to the left handed neutrinos. In fact the third right handed neutrino is not required to create the desired mixing matrix \cite{matheta13,LavLmodel2}. So for models with only two right handed neutrinos the Dirac mass matrix
\begin{equation}\label{favdirac3}
M_{Dirac\;3}=\left(\begin{array}{cc}X&0\\0&X\\0&X\end{array}\right),
\end{equation}
in combination with some $2\times 2$ right handed neutrino mass matrices can give diagonalisation matrices with one zero element.

The first Dirac mass matrix (Eq. \ref{favdirac}) can be created from the transformations
\begin{equation}
X_L=diag(e^{i \alpha_1},e^{i \alpha_2},e^{i \alpha_2}),\: X_{\nu R}=diag(e^{i \alpha_1},e^{i \alpha_2},e^{i \alpha_3}).
\end{equation}
This transformation also ensures that $M_{\ell} M_{\ell}^\dagger$ has a lower right $2\times 2$ block form.
There are 20 different right handed neutrino mass matrices that when combined with this Dirac mass matrix via $M_\nu=-M_{Dirac\;1}M_R^{-1}M_{Dirac\;1}^T$ give $\theta_{13}=0$. These matrices are listed in table \ref{tabmajfav1}.

\begin{table}
\begin{center}
\begin{tabular}{|l|l|l|l|l|l|}
\hline
$\left(\begin{array}{ccc}0&X&X\\X&X&X\\X&X&X\end{array}\right)$&
$\left(\begin{array}{ccc}X&X&X\\X&0&X\\X&X&X\end{array}\right)$&
$\left(\begin{array}{ccc}X&X&X\\X&X&X\\X&X&0\end{array}\right)$&
$\left(\begin{array}{ccc}0&X&X\\X&0&X\\X&X&0\end{array}\right)$&
$\left(\begin{array}{ccc}0&X&X\\X&0&X\\X&X&X\end{array}\right)$
\\
\hline
$\left(\begin{array}{ccc}0&X&X\\X&X&X\\X&X&0\end{array}\right)$&
$\left(\begin{array}{ccc}X&X&X\\X&0&X\\X&X&0\end{array}\right)$&
$\left(\begin{array}{ccc}X&0&X\\0&X&X\\X&X&X\end{array}\right)$&
$\left(\begin{array}{ccc}X&X&0\\X&X&X\\0&X&X\end{array}\right)$&
$\left(\begin{array}{ccc}X&X&X\\X&X&0\\X&0&X\end{array}\right)$
\\
\hline
$\left(\begin{array}{ccc}0&X&X\\X&0&0\\X&0&X\end{array}\right)$&
$\left(\begin{array}{ccc}0&X&0\\X&0&X\\0&X&X\end{array}\right)$&
$\left(\begin{array}{ccc}X&0&X\\0&0&X\\X&X&0\end{array}\right)$&
$\left(\begin{array}{ccc}0&0&X\\0&X&X\\X&X&0\end{array}\right)$&
$\left(\begin{array}{ccc}X&X&0\\X&0&0\\0&0&X\end{array}\right)$
\\
\hline
$\left(\begin{array}{ccc}0&X&0\\X&X&0\\0&0&X\end{array}\right)$&
$\left(\begin{array}{ccc}X&0&X\\0&0&X\\X&X&X\end{array}\right)$&
$\left(\begin{array}{ccc}0&0&X\\0&X&X\\X&X&X\end{array}\right)$&
$\left(\begin{array}{ccc}X&X&0\\X&X&0\\0&0&X\end{array}\right)$&
$\left(\begin{array}{ccc}X&X&X\\X&X&X\\X&X&X\end{array}\right)$
\\
\hline
\end{tabular}
\end{center}
\caption{Right handed neutrino mass matrices that give $\theta_{13}=0$ when combined with the Dirac mass matrix in Eq. \ref{favdirac}.}
\label{tabmajfav1}
\end{table}

The Dirac mass matrix of Eq. \ref{favdirac2} can be generated by the transformations
\begin{equation}
X_L=diag(e^{i \alpha_1},e^{i \alpha_2},e^{i \alpha_2}),\: X_{\nu R}=diag(e^{i \alpha_1},e^{i \alpha_2},e^{i \alpha_1}),
\end{equation}
which again, also ensures that $M_{\ell} M_{\ell}^\dagger$ has a lower right $2\times 2$ block form. Note that the first and the third right handed neutrinos transform in the same way, indicating that the right handed neutrino mass matrices must have the same form after an interchange of the first and third rows and columns. There are three Majorana matrices that have this property and give $\theta_{13}=0$ when coupled to Eq. \ref{favdirac2}:
\begin{equation}\label{favmaj2}
M_R=\left(\begin{array}{ccc}0&X&X\\X&0&X\\X&X&0\end{array}\right),\:
\left(\begin{array}{ccc}X&X&X\\X&X&X\\X&X&X\end{array}\right),\:
\left(\begin{array}{ccc}X&X&X\\X&0&X\\X&X&X\end{array}\right).\:
\end{equation}

The Dirac mass matrix of Eq. \ref{favdirac3}, and a lower right $2\times 2$ block $M_{\ell} M_{\ell}^\dagger$  can be created by the transformations 
\begin{equation}
X_L=diag(e^{i \alpha_1},e^{i \alpha_2},e^{i \alpha_2}),\: X_{\nu R}=diag(e^{i \alpha_1},e^{i \alpha_2}).
\end{equation}
This mass matrix in conjunction with one of the right handed neutrino mass matrices
\begin{equation}\label{2by2maj}
M_R=\left(\begin{array}{cc}X&X\\X&X\end{array}\right),\: \left(\begin{array}{cc}X&X\\X&0\end{array}\right),\textrm{ and } \left(\begin{array}{cc}0&X\\X&X\end{array}\right), 
\end{equation}
generates a mixing matrix with $\theta_{13}=0$.

There are also sets of matrices that produce $\theta_{13}=0$ from the general seesaw mechanism given by $M_\nu=M_L-M_{Dirac}M_R^{-1}M_{Dirac}^T$. However, in all these cases the seesaw part of this matrix ($-M_{Dirac}M_R^{-1}M_{Dirac}^T$) also gives $\theta_{13}=0$, so the addition of the $M_L$ Majorana matrix unnecessarily complicates the model, therefore I have not presented these matrices.

All the matrices listed in this subsection have resultant $M_\nu$ matrices that have no zero elements, but the elements of $M_{\nu}$ are related due to the form of the Dirac and right handed Majorana matrices that they are comprised of. 
Every set of $M_R$ and $M_{Dirac}$ matrices predict $\nu_3$ to be massless giving an inverse hierarchy of neutrino masses.
Using $\delta m^2_{12}=8.3 \times 10^{-5} eV^2$ and $\delta m^2_{23}=2.4 \times 10^{-3} eV^2$ \cite{fog04}, the neutrino masses are $m_3=0 \,eV$, $m_2=0.049 \,eV$, and $m_1=0.058 \,eV$ or $0.040\, eV$.

This section showed that it is possible for models with one Higgs doublet to create a neutrino mass matrix with $\theta_{13}=0$.
 These models all have a neutrino diagonalisation matrix $U_\nu$ with one zero element, and a left handed lepton diagonalisation matrix $U_{\ell L}$ which has a lower right $2\times 2$ block form. Neutrino-triplet Higgs couplings are able to generate mass matrices that give $\theta_{13}=0$, but do not allow the known solar mixing angle and $\delta m^2$.
Models utilising the seesaw mechanism and Higgs singlets can create $\theta_{13}=0$ and give no restrictions on solar or atmospheric mixing. These models can have two or three right handed neutrinos.

\section{The models that give $\theta_{13}=0$ with only one real scalar singlet}\label{favmodels}

\begin{table}
\begin{center}
\begin{tabular}{|l|l|l|l|l|l|l|l|l|}
\hline
&$M_{Bare}$&$\epsilon$&$M_{Dirac}$&$M_\ell$&Symmetry&$X_L=X_{\ell R}$&$X_{\nu R}$&Singlet\\
&&&&&&&&transformation\\
\hline \hline
1&
$\left(\begin{array}{ccc}X&0&0\\0&X&0\\0&0&0\end{array}\right)$&
$\left(\begin{array}{ccc}0&X&0\\X&0&0\\0&0&X\end{array}\right)$&
$\left(\begin{array}{ccc}X&0&0\\0&X&0\\0&X&0\end{array}\right)$&
$\left(\begin{array}{ccc}X&0&0\\0&X&X\\0&X&X\end{array}\right)$&
$Z_4$&
$\left(\begin{array}{ccc}1&0&0\\0&-1&0\\0&0&-1\end{array}\right)$&
$\left(\begin{array}{ccc}1&0&0\\0&-1&0\\0&0&i\end{array}\right)$&
$\chi \rightarrow -\chi $
\\
\hline
2&
$\left(\begin{array}{ccc}X&0&X\\0&X&0\\X&0&X\end{array}\right)$&
$\left(\begin{array}{ccc}0&X&0\\X&0&X\\0&X&0\end{array}\right)$&
$\left(\begin{array}{ccc}X&0&X\\0&X&0\\0&X&0\end{array}\right)$&
$\left(\begin{array}{ccc}X&0&0\\0&X&X\\0&X&X\end{array}\right)$&
$Z_2$&
$\left(\begin{array}{ccc}1&0&0\\0&-1&0\\0&0&-1\end{array}\right)$&
$\left(\begin{array}{ccc}1&0&0\\0&-1&0\\0&0&1\end{array}\right)$&
$\chi \rightarrow -\chi$
\\
\hline
3&
$\left(\begin{array}{ccc}X&0\\0&X\end{array}\right)$&
$\left(\begin{array}{ccc}0&X\\X&0\end{array}\right)$&
$\left(\begin{array}{cc}X&0\\0&X\\0&X\end{array}\right)$&
$\left(\begin{array}{ccc}X&0&0\\0&X&X\\0&X&X\end{array}\right)$&
$Z_2$&
$\left(\begin{array}{ccc}1&0&0\\0&-1&0\\0&0&-1\end{array}\right)$&
$\left(\begin{array}{cc}1&0\\0&-1\end{array}\right)$&
$\chi \rightarrow -\chi $
\\
\hline
\end{tabular}
\end{center}
\caption{Models that give $\theta_{13}=0$ for seesaw neutrinos with couplings to one real  singlet Higgs.}
\label{tabseesaw}
\end{table}

None of the mass matrices listed in Sec. \ref{onezeronu} can be created with models with only one doublet and no other Higgs fields. The minimal additional field needed to produce any of these sets of matrices is one real scalar singlet.
Table \ref{tabseesaw} lists the mass matrices that can be created using one real scalar singlet, and the symmetry and transformations used. There is one model for each of the three Dirac mass matrices in Eq.s \ref{favdirac}-\ref{favdirac3}. The right handed neutrinos gain mass through a bare mass term, as well as coupling to the singlet, giving a right handed Majorana mass matrix of $M_R=M_{Bare}+\epsilon \langle \chi \rangle$.

The mass matrices $M_{\ell}$, $M_{Dirac}$ and $M_R$ in model 3 are the same as the model suggested by Grimus and Lavoura \cite{LavLmodel2}. Their model was generated using an $L_e-L_\mu-L_\tau$ symmetry, which was explicitly broken to generate the diagonal elements in the right handed Majorana mass matrix. The model listed here instead generates the extra elements in $M_R$ using the VEV of a real Higgs singlet as described in the table, but if the $L_e-L_\mu-L_\tau$ symmetry is used, a complex Higgs singlet with an $L_e-L_\mu-L_\tau$ charge of $2$ will also create the required mass matrices. 

These three models demonstrate that it is possible for a model with family symmetry to reduce the number of free parameters in the lepton masses and mixing matrix. However, if the total number of free parameters is more in the model than in the SM with neutrino masses, it is difficult to justify the addition of one new Higgs field, so the number of free variables in these models should be calculated, and compared to models with the same mass generation mechanism, but with no family symmetry.

The number of physical parameters in the lepton sector with three right handed neutrinos, no family symmetry, and no Higgs singlets is twenty-one, corresponding to nine masses, six independent angles, and six phases \cite{santamaria,jenkins}. These variables arise from the two Dirac mass matrices and the right handed Majorana mass matrix.
A basis where the number of free variables is evident is the weak basis where $M_\ell$ and $M_R$ are diagonal and real. These two mass matrices contribute three parameters each. 
This choice of basis completely defines the right handed neutrino basis, but the left handed leptons can still change basis via a multiplication of a diagonal unitary matrix with three independent phases. These three phases can eliminate three free variables in $M_{Dirac}$, ensuring that three of the nine parameters are real. The total number of parameters is twenty-one: three from $M_\ell$, three from $M_R$ and fifteen from $M_{Dirac}$. 

The seesaw mechanism with two right handed neutrinos has fourteen physical parameters corresponding to seven masses, four independent angles, and three phases. The diagonal $M_\ell$ has three parameters, the diagonal $M_R$ has two parameters and $M_{Dirac}$ which is a $3\times 2$ matrix has nine parameters.

The number of variables in the mass matrices listed in table \ref{tabseesaw} can also be reduced by choosing the basis where $M_\ell$ and $M_{Bare}$ are real and diagonal, and by using any extra freedom in the basis choice to reduce the number of phases in $M_{Dirac}$ and $\epsilon$. The mass matrices for the three models in this basis are listed in table \ref{tabbasis}. 

The first model has three variables each in $M_\ell$, $\epsilon$, and $M_{Dirac}$ and two variables in $M_{Bare}$, giving a total of eleven free parameters associated with the leptons, ten fewer parameters than the seesaw model with three right handed neutrinos and no family symmetry.

The second model has three parameters each in $M_\ell$ and $M_{Bare}$, four parameters in $\epsilon$ and five parameters in $M_{Dirac}$, giving fifteen parameters in the lepton mass matrices, six fewer than in the seesaw model with three right handed neutrinos.

The third model has two variables each in $\epsilon$ and $M_{Bare}$, and three each in $M_\ell$ and $M_{Dirac}$, giving ten variables, four fewer than in lepton mass matrices in the seesaw model with two right handed neutrinos.

To see whether these models have fewer parameters overall, consider the Higgs potential for these models.
The most general Higgs potential for one Higgs doublet, and one Higgs singlet that transforms via $\chi\rightarrow -\chi$ is  
\begin{equation}\label{higgspot}
V= c_1 \phi^\dagger \phi + c_2 (\phi^\dagger \phi)^2 + c_3 \chi^2 + c_4 \chi^4 + c_5 \chi^2  \phi^\dagger \phi,  
\end{equation}
where $c_{1-5}$, are the five free variables of the Higgs potential, three more than a Higgs potential with only one SM doublet. So the first model has seven fewer parameters than the seesaw model with three right handed neutrinos and no family symmetry, the second model has three fewer parameters, and the third model has one fewer parameter than the seesaw model with two right handed neutrinos.

All three models have fewer free variables than models with no family symmetry and the same number of right handed neutrinos.

\begin{table}
\begin{center}
\begin{tabular}{|l|l|l|l|l|}
\hline
&$M_{Bare}$&$\epsilon$&$M_{Dirac}$&$M_\ell$\\
\hline \hline
1&
$\left(\begin{array}{ccc}A&0&0\\0&B&0\\0&0&0\end{array}\right)$&
$\left(\begin{array}{ccc}0&D e^{i \delta}&0\\D e^{i \delta}&0&0\\0&0&E\end{array}\right)$&
$\left(\begin{array}{ccc}a_1&0&0\\0&a_2&0\\0&a_3&0\end{array}\right)$&
$\left(\begin{array}{ccc}m_e&0&0\\0&m_\mu&0\\0&0&m_\tau\end{array}\right)$
\\
\hline
2&
$\left(\begin{array}{ccc}A&0&0\\0&B&0\\0&0&C\end{array}\right)$&
$\left(\begin{array}{ccc}0&D e^{i \delta}&0\\D e^{i \delta}&0&E e^{i \gamma}\\0&E e^{i \gamma}&0\end{array}\right)$&
$\left(\begin{array}{ccc}a_1&0&a_4 e^{i \alpha}\\0&a_2&0\\0&a_3&0\end{array}\right)$&
$\left(\begin{array}{ccc}m_e&0&0\\0&m_\mu&0\\0&0&m_\tau\end{array}\right)$
\\
\hline
3&
$\left(\begin{array}{ccc}A&0\\0&B\end{array}\right)$&
$\left(\begin{array}{ccc}0&D e^{i \delta}\\D e^{i \delta}&0\end{array}\right)$&
$\left(\begin{array}{cc}a_1&0\\0&a_2\\0&a_3\end{array}\right)$&
$\left(\begin{array}{ccc}m_e&0&0\\0&m_\mu&0\\0&0&m_\tau\end{array}\right)$
\\
\hline
\end{tabular}
\end{center}
\caption{Models of table \ref{tabseesaw} in the basis where $M_\ell$ and $M_{Bare}$ are real and diagonal. All parameters in the mass matrices are real.}
\label{tabbasis}
\end{table}

\section{Mass matrix renormalisation}\label{rges}

The predictions of $\theta_{13}=0$ and $m_{\nu 3}=0$ in the models in Sec. \ref{onezeronu} hold at the seesaw scale, so it is useful to analyse the effect of the renormalisation group running to ascertain whether these predictions hold at a low energy scale. 

Under one-loop radiative corrections the Majorana neutrino mass matrix in the charged lepton mass basis at a high scale $M_{\nu mass}$ will evolve to a low scale according to   
\begin{equation}
M^{'}_{\nu mass}= R M_{\nu mass} R, 
\end{equation}
where $R=diag(R_e, R_\mu, R_\tau)$, and $R_e$, $R_\mu$, and $R_\tau$ are numbers relating to the charged lepton Yukawa couplings and the energy scales \cite{rgesmirnov,rgemunich}.

The transformation $M_{\nu mass}= U^\dagger_{\ell L} M_\nu U^*_{\ell L}$ changes a Majorana mass matrix into the charged lepton mass basis, where $U_{\ell L}$ is the diagonalisation matrix for the charged lepton. In the basis where the transformation matrices are diagonal, all the models presented give
\begin{equation}
U_{\ell L}=\left(\begin{array}{ccc}1&0&0\\0&\cos \alpha & \sin \alpha \\0&-\sin \alpha & \cos \alpha \end{array}\right).
\end{equation}

For the matrices of Eqs. \ref{maj1} and \ref{maj2}, where the neutrinos gain mass via a triplet Higgs coupling, the pattern of zeros and unknown elements in the neutrino mass matrix is unchanged by moving to the charged lepton mass basis. The one-loop renormalisation group running does not make zero elements non-zero,  only the non-zero elements are altered. However since the non-zero elements are in any case unrestricted by the symmetry the renormalisation group running does not change any of the predictions of these two matrices.

For the cases where the neutrino mass matrix arises from the seesaw mechanism, the change to the charged lepton mass basis, and the renormalisation group running, can be expressed as a change in the Dirac mass matrix: $M_{Dirac\;mass}^{'}= R U^\dagger_{\ell L}M_{Dirac}$.

The Dirac mass matrix in Eq. \ref{favdirac} has the form
\begin{equation}
M_{Dirac}=\left(\begin{array}{ccc}a_1&0&0\\0&a_2&0\\0&a_3&0\end{array}\right),
\end{equation}
so the renormalised Dirac mass matrix in the charged lepton mass basis is
\begin{equation}
M^{'}_{Dirac\;mass}=
\left(\begin{array}{ccc}a_1^{'}&0&0\\0&a_2^{'}&0\\0&a_3^{'}&0\end{array}\right),
\end{equation}
where $a_1^{'}=a_1 R_e$, $a_2^{'}=R_\mu (a_2 \cos \alpha - a_3 \sin \alpha )$ and $a_3^{'}=R_\tau (a_2 \sin \alpha + a_3 \cos \alpha  )$.
Again, since the texture of the Dirac mass matrix is the same before and after renormalisation group running, and the non-zero entries in $ M^{'}_{Dirac\;mass}$ are free parameters, the predictions are not changed by the one-loop renormalisation group running. 
The Dirac mass matrix of Eq. \ref{favdirac2} in table \ref{tabseesaw} is also unchanged by the renormalisation group. 

Therefore, all of the models that have one Higgs doublet and predict $\theta_{13}=0$ still predict this after one-loop renormalisation group running, and still predict that $\nu_3$ is massless. The predictions of mass matrices due to triplet Higgs coupling (Eq.s \ref{maj1} and \ref{maj2}) also remain the same, so the solar mixing produced by these matrices remain outside experimental bounds.    
If a non-zero $\theta_{13}$ is observed, or if the absolute scale of the neutrino mass is found to be larger than $0.06 eV$ the three models of table \ref{tabseesaw} will be falsifed.

\section{Conclusion}

Family symmetries may provide a way of explaining the best-fit form of the neutrino mixing matrix, however, it was shown that Abelian family symmetries can only explain one aspect of the mixing matrix form: the smallness of $\theta_{13}$. The minimal model that predicts $\theta_{13}=0$, and has allowed values for the other mixing angles involves one real singlet Higgs, a $Z_2$ symmetry and two right handed neutrinos, although a third right handed neutrino can also be accommodated. This model predicts an inverse hierarchy, and $\theta_{13}$ remains zero under one-loop renormalisation group running. 
The free variables in these models were counted in order to ascertain whether this model has advantages over the SM with neutrino masses. The SM with two right handed neutrinos has sixteen parameters in the lepton and Higgs sectors, and with three right handed neutrinos there are twenty-three parameters. The $Z_2$ model with two right handed neutrinos and a Higgs singlet has one fewer free variable than the case with no family symmetry, and the simplest model with three right handed neutrinos has seven fewer parameters than its equivalent model with no family symmetry. So these models not only ensure $\theta_{13}=0$, they also reduce the number of free variables overall.

To explain aspects of the mixing matrix other than the smallness of $\theta_{13}$, a non-Abelian symmetry will be required, however in many simple models a non-Abelian symmetry guarantees unrealistic lepton masses. To produce a viable non-Abelian symmetry model either more than one Higgs doublet is required, or the neutrinos must gain mass through two different mechanisms: The seesaw mechanism and coupling to a triplet Higgs or via a dimension-5 operator. So either we have to be content with family symmetry models predicting only a small aspect of the mixing matrix, or we have to accept the consequences of more complex models. 

\begin{acknowledgements}
I would like to thank R. Volkas for many useful discussions and assistance with the draft. This work was supported in part by The University of Melbourne, and in part by the 
Australian Research Council.
\end{acknowledgements}

\appendix

\section{Equivalent representations of the lepton transformations}\label{app}

This appendix shows that two different, but equivalent, representations for the lepton transformation matrices give exactly the same restrictions on the lepton mixing matrix. It follows the structure of Appendix A in reference \cite{nogo2} but generalises that result. The method is to find the restrictions on the mass matrices due to the transformations, and since the transformation matrices are related via a similarity transformation, the restrictions on the mass matrices from both sets of transformations are also related. The relationship between the diagonalisation matrices from the two representations are then found, and the mixing matrices are compared.

Two equivalent representations for the fermion transformations, $A_L,A_{\ell R},A_{\nu R}$ and $B_L,B_{\ell R},B_{\nu R}$, are related by 
\begin{eqnarray}
B_L&=&S_L^\dagger A_L S_L, \\
B_{\ell R}&=&S_{\ell R}^\dagger A_{\ell R} S_{\ell R},\\
B_{\nu R}&=&S_{\nu R}^\dagger A_{\nu R} S_{\nu R}.
\end{eqnarray}
where $S_L,S_{\ell R}$ and $S_{\nu R}$ can be any $3\times 3$ unitary matrices. 
The Higgs doublets, singlets and triplets transform respectively as
\begin{equation}
\Phi \rightarrow V_\Phi \Phi, \: \chi \rightarrow V_\chi S, \: \Delta \rightarrow V_\Delta \Delta .
\end{equation}

Reference \cite{nogo2} demonstrated that the charged lepton mass matrix derived from the $A$ transformations, $M_{\ell A}$, is related to the charged lepton mass matrix from the $B$ transformation, $M_{\ell B}$. $M_{\ell A}$ has the same restrictions from the symmetry as $S_L M_{\ell B} S_{\ell R}^\dagger$, so the two matrices have equivalent forms. Diagonalising these two mass matrices shows that $U_{\ell L}^B$ has the same restrictions from the symmetry as $S_L^\dagger U_{\ell L}^A$, where  $U_{\ell L}^{A,B}$ are the left diagonalisation matrices for $M_{\ell A}$ and $M_{\ell B}$ respectively.
Similarly, the Dirac neutrino mass matrix $M_{Dirac\: A}$ has the same restrictions due to the symmetry as $S_L M_{Dirac\: B} S_{\nu R}^\dagger$.

Reference \cite{nogo2} also shows that neutrino mass matrices arising from Dimension-5 operators or the seesaw mechanism with no Higgs singlets have similar properties.  $M_{\nu A}$ has the same restrictions as $S_L M_{\nu B} S_L^T$, so by setting $M_{\nu A}=S_L M_{\nu B} S_L^T$ and diagonalising, the diagonalisation matrices can be shown to be related by  $U_{\nu}^B=S_L^\dagger U_{\nu}^A$. Putting the neutrino and charged lepton diagonalisation matrices together demonstrates that the mixing matrices are the same:
\begin{eqnarray}
U_{\mathrm{MNS}}^B&=&U_{\ell L}^{B \dagger} U_{\nu}^B= U_{\ell L}^{A \dagger} S_L S_L^\dagger U_{\nu}^A\nonumber\\
&=&U_{\mathrm{MNS}}^A.\label{mnssame}
\end{eqnarray}

Next consider neutrinos that gain mass by coupling to a Higgs triplet.
In this case, the neutrino mass matrix is given by $M_\nu = \lambda_i \langle \Delta^0_i \rangle$.
The coupling matrices $\lambda_i$ are constrained by the $A$ transformation according to
\begin{equation}
\lambda_{A i}=A_L^\dagger V_{\Delta ij} \lambda_{A j} A_L^*,
\end{equation} 
and are constrained by the $B$ transformation according to
\begin{eqnarray}
\lambda_{B i}&=&B_L^\dagger V_{\Delta ij} \lambda_{B j} B_L^*\\
&=&S_L^\dagger A_L^\dagger S_L   V_{\Delta ij} \lambda_{B j} S_L^T A_L^* S_L^*.
\end{eqnarray} 
Rearranging shows that
\begin{equation}
(S_L\lambda_{B i}S_L^T)=A_L^\dagger   V_{\Delta ij} (S_L \lambda_{B j} S_L^T) A_L^*,
\end{equation} 
demonstrating that
$S_L \lambda_{B i} S_L^T$ has the same restrictions due to the symmetry as $\lambda_{A i}$, so they can be equated. 
Forming the mass matrix for $A$ shows
\begin{eqnarray}
M_{\nu A}& =& \lambda_{A i} \langle \Delta^0_i \rangle\\
&=&S_L \lambda_{B i} \langle \Delta^0_i \rangle S_L^T\\
&=&S_L M_{\nu B} S_L^T,
\end{eqnarray}
demonstrating that $M_{\nu A}$ has exactly the same restrictions due to the symmetry as 
$S_L M_{\nu B} S_L^T$. Diagonalising shows that the diagonalisation matrices must be related by $U_{\nu}^B=S_L^\dagger U_{\nu}^A$, and using Eq. \ref{mnssame} the mixing matrices from the $A$ and $B$ transformations can be shown to have identical forms.

When singlet Higgs fields are present, and the right handed neutrinos gain mass via coupling to the singlet Higgs and a bare mass term, the right handed Majorana mass matrix is given by $M_R=\epsilon_i \langle \chi_i \rangle + M_{Bare}$. 
The coupling matrices $\epsilon_i$ are restricted by the $B$ transformations by
\begin{eqnarray}
\epsilon_{B i}&=&B_{\nu R}^T V_{\chi ij} \epsilon_{B j}B_{\nu R} \\
&=&S_{\nu R}^T A_{\nu R}^T S_{\nu R}^* V_{\chi ij} \epsilon_{B j} S_{\nu R}^\dagger A_{\nu R} S_{\nu R}.  
\end{eqnarray}
Rearranging this equation shows 
\begin{equation}
(S_{\nu R}^* \epsilon_{B i} S_{\nu R}^\dagger)= A_{\nu R}^T V_{\chi\;ij} (S_{\nu R}^* \epsilon_{B j} S_{\nu R}^\dagger) A_{\nu R},
\end{equation}
demonstrating that  $S_{\nu R}^* \epsilon_{B i} S_{\nu R}^\dagger$ has the same restrictions as $\epsilon_{A i}$. 
Similarly, the bare mass matrix from the $B$ transformation, $M_{Bare\; B}$, is related to that of the $A$ transformation by $M_{Bare\; A}=S_{\nu R}^* M_{Bare\;B} S_{\nu R}^\dagger$. So the total right handed mass matrix from the $A$ transformation is 
\begin{eqnarray}
M_{R\;A}&=&\epsilon_{A\;i} \langle \chi_i \rangle + M_{Bare\;A}\\
&=&S_{\nu R}^* (\epsilon_{B\;i} \langle \chi_i \rangle + M_{Bare\;B})S_{\nu R}^\dagger\\
&=&S_{\nu R}^* M_{R\;B} S_{\nu R}^\dagger.
\end{eqnarray}
Since the two Dirac mass matrices are related by $M_{Dirac\: A}=S_L M_{Dirac\: B} S_{\nu R}^\dagger$, the resultant left handed neutrino mass matrix is
\begin{eqnarray}
M_{\nu A}&=&M_{Dirac\: A} M_{R\;A}^{-1}M_{Dirac\: A}^T\\
&=&S_L M_{Dirac\: B} S_{\nu R}^\dagger S_{\nu R} M_{R\;B}^{-1} S_{\nu R}^T S_L^* M_{Dirac\: B}^T S_{\nu R}^T\\
&=&S_L M_{\nu B} S_{\nu R}^T.
\end{eqnarray} 
Again, the relationship between the mass matrices means that $U_{\nu}^B=S_L^\dagger U_{\nu}^A$, and using Eq. \ref{mnssame}, the mixing matrix is the same for both the $A$ and $B$ transformations. 

All methods of gaining Majorana neutrino masses show the relationship $M_{\nu A}=S_L M_{\nu B} S_L^T$, so a model that has neutrino mass gained from a number of these methods will also display this relationship between the mass matrices from different, but equivalent representations. This means that for models with any combination of the methods shown above, two equivalent representations will still give identical results. 

\section{Non-Abelian right handed neutrino transformations}\label{appnonAb}

The following shows that in models with one Higgs doublet, left handed neutrinos transforming via an Abelian group, and right handed neutrinos transforming via a non-Abelian group always result in $M_{Dirac}$ having at least two zero eigenvalues. 

For the right handed neutrinos to transform via a non-Abelian group there needs to be at least two transformation matrices that do not commute, two of which I label $X_{\nu R 1}$ and $X_{\nu R 2}$. The left handed transformations that act in conjunction with these right handed transformations are labelled $X_{L 1}$ and $X_{L 2}$. 
The Dirac mass matrix must obey the symmetry restrictions
\begin{equation}
M_{Dirac}=X_{L 1}^\dagger M_{Dirac} X_{\nu R 1}=X_{L 2}^\dagger M_{Dirac} X_{\nu R 2}.
\end{equation}
If the basis where $X_{L 1}$,  $X_{L 2}$ and $X_{\nu R 1}$ are diagonal is chosen, the first transformation acts to enforce some texture zeros on the mass matrix. The only mass matrices that can arise from such a transformation, and have at least two non-zero eigenvalues are
\begin{eqnarray}\label{twoevals}
\left(\begin{array}{ccc}0&0&0\\0&a_2&0\\0&0&a_1\end{array}\right),& 
\left(\begin{array}{ccc}0&0&0\\a_3&0&0\\0&a_2&a_1\end{array}\right),& 
\left(\begin{array}{ccc}0&a_3&0\\0&0&a_2\\0&0&a_1\end{array}\right),\\
\left(\begin{array}{ccc}0&0&0\\0&a_4&a_3\\0&a_2&a_1\end{array}\right),& 
\left(\begin{array}{ccc}0&0&a_4\\0&0&a_3\\a_2&a_1&0\end{array}\right),& 
\left(\begin{array}{ccc}a_5&0&0\\0&a_4&a_3\\0&a_2&a_1\end{array}\right),\nonumber\\ 
\left(\begin{array}{ccc}0&a_6&a_5\\0&a_4&a_3\\0&a_2&a_1\end{array}\right),& 
\left(\begin{array}{ccc}a_3&0&0\\0&a_2&0\\0&0&a_1\end{array}\right),\nonumber
\end{eqnarray} 
and row and column interchanges of these matrices.

Applying the second transformation to these already constrained matrices will always reduce the number of non-zero eigenvalues to zero or one. 
To illustrate this just consider the first matrix in Eq. \ref{twoevals}. 
To create this matrix the first transformation matrices have the form
\begin{eqnarray}
X_{L 1}&=&diag(e^{i \alpha_1},e^{i \alpha_2},e^{i \alpha_3}),\\
X_{\nu R 1}&=&diag(e^{i \alpha_4},e^{i \alpha_2},e^{i \alpha_3}),
\end{eqnarray} 
and the action of the second transformation is
\begin{eqnarray}
M_{Dirac}&=&X_{L 2}^\dagger M_{Dirac} X_{\nu R 2}\\
&=&\left(\begin{array}{ccc}e^{-i \beta_1}&0&0\\0&e^{-i \beta_2}&0\\0&0&e^{-i \beta_3}\end{array}\right)\left(\begin{array}{ccc}0&0&0\\0&a_2&0\\0&0&a_1\end{array}\right) U,
\end{eqnarray}
where $U=X_{\nu R 2}$.
The restrictions on elements 22 and 33 of matrix $M_{Dirac}$ are 
\begin{equation}
a_1=e^{-i \beta_2} a_1 U_{22}, \: a_2=e^{-i \beta_3} a_1 U_{33}.
\end{equation}
If both these elements are to remain non-zero, $e^{-i \beta_2} U_{22}=e^{-i \beta_3}U_{33}=1$, therefore $|U_{22}|=|U_{33}|=1$, and the unitary matrix must be diagonal. A diagonal $X_{\nu R 2}$ will commute with $X_{\nu R1}$, meaning that the right handed neutrinos actually transform via an Abelian group. 
This is the same for all of the other matrices in Eq. \ref{twoevals} --  either the mass matrix has two or three zero eigenvalues, or the right handed neutrinos transform according to an Abelian group.

\bibliography{paper}

\begin{thebibliography}{49}
\expandafter\ifx\csname natexlab\endcsname\relax\def\natexlab#1{#1}\fi
\expandafter\ifx\csname bibnamefont\endcsname\relax
  \def\bibnamefont#1{#1}\fi
\expandafter\ifx\csname bibfnamefont\endcsname\relax
  \def\bibfnamefont#1{#1}\fi
\expandafter\ifx\csname citenamefont\endcsname\relax
  \def\citenamefont#1{#1}\fi
\expandafter\ifx\csname url\endcsname\relax
  \def\url#1{\texttt{#1}}\fi
\expandafter\ifx\csname urlprefix\endcsname\relax\def\urlprefix{URL }\fi
\providecommand{\bibinfo}[2]{#2}
\providecommand{\eprint}[2][]{\url{#2}}

\bibitem[{\citenamefont{Minkowski}(1977)}]{seesaw1}
\bibinfo{author}{\bibfnamefont{P.}~\bibnamefont{Minkowski}},
  \bibinfo{journal}{Phys. Lett.} \textbf{\bibinfo{volume}{B67}},
  \bibinfo{pages}{421} (\bibinfo{year}{1977}).

\bibitem[{\citenamefont{Glashow}(1979)}]{seesaw2}
\bibinfo{author}{\bibfnamefont{S.~L.} \bibnamefont{Glashow}}, in
  \emph{\bibinfo{booktitle}{Quarks and leptons, proceedings of the advanced
  study institute}} (\bibinfo{year}{1979}).

\bibitem[{\citenamefont{Gell-Mann et~al.}(1979)\citenamefont{Gell-Mann, Ramond,
  and Slansky}}]{seesaw3}
\bibinfo{author}{\bibfnamefont{M.}~\bibnamefont{Gell-Mann}},
  \bibinfo{author}{\bibfnamefont{P.}~\bibnamefont{Ramond}}, \bibnamefont{and}
  \bibinfo{author}{\bibfnamefont{R.}~\bibnamefont{Slansky}}, in
  \emph{\bibinfo{booktitle}{Supergravity}} (\bibinfo{year}{1979}).

\bibitem[{\citenamefont{Yanagida}(1979)}]{seesaw4}
\bibinfo{author}{\bibfnamefont{T.}~\bibnamefont{Yanagida}}, in
  \emph{\bibinfo{booktitle}{Proc. of the workshop on Unified Theory and Baryon
  number in the Universe}} (\bibinfo{year}{1979}).

\bibitem[{\citenamefont{Mohapatra and Senjanovic}(1980)}]{seesaw5}
\bibinfo{author}{\bibfnamefont{R.~N.} \bibnamefont{Mohapatra}}
  \bibnamefont{and}
  \bibinfo{author}{\bibfnamefont{G.}~\bibnamefont{Senjanovic}},
  \bibinfo{journal}{Phys. Rev. Lett.} \textbf{\bibinfo{volume}{44}},
  \bibinfo{pages}{912} (\bibinfo{year}{1980}).

\bibitem[{\citenamefont{Santamaria}(1993)}]{santamaria}
\bibinfo{author}{\bibfnamefont{A.}~\bibnamefont{Santamaria}},
  \bibinfo{journal}{Phys. Lett.} \textbf{\bibinfo{volume}{B305}},
  \bibinfo{pages}{90} (\bibinfo{year}{1993}), \eprint{hep-ph/9302301}.

\bibitem[{\citenamefont{Broncano et~al.}(2003)\citenamefont{Broncano, Gavela,
  and Jenkins}}]{jenkins}
\bibinfo{author}{\bibfnamefont{A.}~\bibnamefont{Broncano}},
  \bibinfo{author}{\bibfnamefont{M.~B.} \bibnamefont{Gavela}},
  \bibnamefont{and} \bibinfo{author}{\bibfnamefont{E.}~\bibnamefont{Jenkins}},
  \bibinfo{journal}{Phys. Lett.} \textbf{\bibinfo{volume}{B552}},
  \bibinfo{pages}{177} (\bibinfo{year}{2003}), \eprint{hep-ph/0210271}.

\bibitem[{\citenamefont{Fukuda et~al.}(1999)}]{SKatm2}
\bibinfo{author}{\bibfnamefont{Y.}~\bibnamefont{Fukuda}} \bibnamefont{et~al.}
  (\bibinfo{collaboration}{Super-Kamiokande}), \bibinfo{journal}{Phys. Lett.}
  \textbf{\bibinfo{volume}{B467}}, \bibinfo{pages}{185} (\bibinfo{year}{1999}),
  \eprint{hep-ex/9908049}.

\bibitem[{\citenamefont{Fukuda et~al.}(1994)}]{KAMatm}
\bibinfo{author}{\bibfnamefont{Y.}~\bibnamefont{Fukuda}} \bibnamefont{et~al.}
  (\bibinfo{collaboration}{Kamiokande}), \bibinfo{journal}{Phys. Lett.}
  \textbf{\bibinfo{volume}{B335}}, \bibinfo{pages}{237} (\bibinfo{year}{1994}).

\bibitem[{\citenamefont{Becker-Szendy et~al.}(1995)}]{IMBatm}
\bibinfo{author}{\bibfnamefont{R.}~\bibnamefont{Becker-Szendy}}
  \bibnamefont{et~al.}, \bibinfo{journal}{Nucl. Phys. Proc. Suppl.}
  \textbf{\bibinfo{volume}{38}}, \bibinfo{pages}{331} (\bibinfo{year}{1995}).

\bibitem[{\citenamefont{Sanchez et~al.}(2003)}]{Soudanatmnew}
\bibinfo{author}{\bibfnamefont{M.}~\bibnamefont{Sanchez}} \bibnamefont{et~al.}
  (\bibinfo{collaboration}{Soudan 2}), \bibinfo{journal}{Phys. Rev.}
  \textbf{\bibinfo{volume}{D68}}, \bibinfo{pages}{113004}
  (\bibinfo{year}{2003}), \eprint{hep-ex/0307069}.

\bibitem[{\citenamefont{Ambrosio et~al.}(2001)}]{MACROatm}
\bibinfo{author}{\bibfnamefont{M.}~\bibnamefont{Ambrosio}} \bibnamefont{et~al.}
  (\bibinfo{collaboration}{MACRO}), \bibinfo{journal}{Phys. Lett.}
  \textbf{\bibinfo{volume}{B517}}, \bibinfo{pages}{59} (\bibinfo{year}{2001}),
  \eprint{hep-ex/0106049}.

\bibitem[{\citenamefont{Apollonio et~al.}(1999)}]{CHOOZ}
\bibinfo{author}{\bibfnamefont{M.}~\bibnamefont{Apollonio}}
  \bibnamefont{et~al.} (\bibinfo{collaboration}{CHOOZ}),
  \bibinfo{journal}{Phys. Lett.} \textbf{\bibinfo{volume}{B466}},
  \bibinfo{pages}{415} (\bibinfo{year}{1999}), \eprint{hep-ex/9907037}.

\bibitem[{\citenamefont{Ahmed et~al.}(2004)}]{SNOsolnew}
\bibinfo{author}{\bibfnamefont{S.~N.} \bibnamefont{Ahmed}} \bibnamefont{et~al.}
  (\bibinfo{collaboration}{SNO}), \bibinfo{journal}{Phys. Rev. Lett.}
  \textbf{\bibinfo{volume}{92}}, \bibinfo{pages}{181301}
  (\bibinfo{year}{2004}), \eprint{nucl-ex/0309004}.

\bibitem[{\citenamefont{Ahn et~al.}(2003)}]{K2Ksol}
\bibinfo{author}{\bibfnamefont{M.~H.} \bibnamefont{Ahn}} \bibnamefont{et~al.}
  (\bibinfo{collaboration}{K2K}), \bibinfo{journal}{Phys. Rev. Lett.}
  \textbf{\bibinfo{volume}{90}}, \bibinfo{pages}{041801}
  (\bibinfo{year}{2003}), \eprint{hep-ex/0212007}.

\bibitem[{\citenamefont{Smy et~al.}(2004)}]{SKsol2}
\bibinfo{author}{\bibfnamefont{M.~B.} \bibnamefont{Smy}} \bibnamefont{et~al.}
  (\bibinfo{collaboration}{Super-Kamiokande}), \bibinfo{journal}{Phys. Rev.}
  \textbf{\bibinfo{volume}{D69}}, \bibinfo{pages}{011104}
  (\bibinfo{year}{2004}), \eprint{hep-ex/0309011}.

\bibitem[{\citenamefont{Abdurashitov et~al.}(2002)}]{SAGEsol}
\bibinfo{author}{\bibfnamefont{J.~N.} \bibnamefont{Abdurashitov}}
  \bibnamefont{et~al.} (\bibinfo{collaboration}{SAGE}), \bibinfo{journal}{J.
  Exp. Theor. Phys.} \textbf{\bibinfo{volume}{95}}, \bibinfo{pages}{181}
  (\bibinfo{year}{2002}).

\bibitem[{\citenamefont{Cleveland et~al.}(1998)}]{HOMESTAKEsol}
\bibinfo{author}{\bibfnamefont{B.~T.} \bibnamefont{Cleveland}}
  \bibnamefont{et~al.}, \bibinfo{journal}{Astrophys. J.}
  \textbf{\bibinfo{volume}{496}}, \bibinfo{pages}{505} (\bibinfo{year}{1998}).

\bibitem[{\citenamefont{Hampel et~al.}(1999)}]{GALLEXsol}
\bibinfo{author}{\bibfnamefont{W.}~\bibnamefont{Hampel}} \bibnamefont{et~al.}
  (\bibinfo{collaboration}{GALLEX}), \bibinfo{journal}{Phys. Lett.}
  \textbf{\bibinfo{volume}{B447}}, \bibinfo{pages}{127} (\bibinfo{year}{1999}).

\bibitem[{\citenamefont{Cattadori}(2002)}]{GNOsol}
\bibinfo{author}{\bibfnamefont{C.~M.} \bibnamefont{Cattadori}}
  (\bibinfo{collaboration}{GNO}), \bibinfo{journal}{Nucl. Phys. Proc. Suppl.}
  \textbf{\bibinfo{volume}{110}}, \bibinfo{pages}{311} (\bibinfo{year}{2002}).

\bibitem[{\citenamefont{Harrison et~al.}(2002)\citenamefont{Harrison, Perkins,
  and Scott}}]{hps1}
\bibinfo{author}{\bibfnamefont{P.~F.} \bibnamefont{Harrison}},
  \bibinfo{author}{\bibfnamefont{D.~H.} \bibnamefont{Perkins}},
  \bibnamefont{and} \bibinfo{author}{\bibfnamefont{W.~G.} \bibnamefont{Scott}},
  \bibinfo{journal}{Phys. Lett.} \textbf{\bibinfo{volume}{B530}},
  \bibinfo{pages}{167} (\bibinfo{year}{2002}), \eprint{hep-ph/0202074}.

\bibitem[{\citenamefont{He and Zee}(2003)}]{zeehe1}
\bibinfo{author}{\bibfnamefont{X.~G.} \bibnamefont{He}} \bibnamefont{and}
  \bibinfo{author}{\bibfnamefont{A.}~\bibnamefont{Zee}},
  \bibinfo{journal}{Phys. Lett.} \textbf{\bibinfo{volume}{B560}},
  \bibinfo{pages}{87} (\bibinfo{year}{2003}), \eprint{hep-ph/0301092}.

\bibitem[{\citenamefont{Kubo et~al.}(2003)\citenamefont{Kubo, Mondragon,
  Mondragon, and Rodriguez-Jauregui}}]{kubo}
\bibinfo{author}{\bibfnamefont{J.}~\bibnamefont{Kubo}},
  \bibinfo{author}{\bibfnamefont{A.}~\bibnamefont{Mondragon}},
  \bibinfo{author}{\bibfnamefont{M.}~\bibnamefont{Mondragon}},
  \bibnamefont{and}
  \bibinfo{author}{\bibfnamefont{E.}~\bibnamefont{Rodriguez-Jauregui}},
  \bibinfo{journal}{Prog. Theor. Phys.} \textbf{\bibinfo{volume}{109}},
  \bibinfo{pages}{795} (\bibinfo{year}{2003}), \eprint{hep-ph/0302196}.

\bibitem[{\citenamefont{Kubo et~al.}(2004)\citenamefont{Kubo, Okada, and
  Sakamaki}}]{kubohiggspot}
\bibinfo{author}{\bibfnamefont{J.}~\bibnamefont{Kubo}},
  \bibinfo{author}{\bibfnamefont{H.}~\bibnamefont{Okada}}, \bibnamefont{and}
  \bibinfo{author}{\bibfnamefont{F.}~\bibnamefont{Sakamaki}},
  \bibinfo{journal}{Phys. Rev.} \textbf{\bibinfo{volume}{D70}},
  \bibinfo{pages}{036007} (\bibinfo{year}{2004}), \eprint{hep-ph/0402089}.

\bibitem[{\citenamefont{Grimus et~al.}(2004{\natexlab{a}})\citenamefont{Grimus,
  Joshipura, Kaneko, Lavoura, and Tanimoto}}]{grimfav}
\bibinfo{author}{\bibfnamefont{W.}~\bibnamefont{Grimus}},
  \bibinfo{author}{\bibfnamefont{A.~S.} \bibnamefont{Joshipura}},
  \bibinfo{author}{\bibfnamefont{S.}~\bibnamefont{Kaneko}},
  \bibinfo{author}{\bibfnamefont{L.}~\bibnamefont{Lavoura}}, \bibnamefont{and}
  \bibinfo{author}{\bibfnamefont{M.}~\bibnamefont{Tanimoto}},
  \bibinfo{journal}{JHEP} \textbf{\bibinfo{volume}{07}}, \bibinfo{pages}{078}
  (\bibinfo{year}{2004}{\natexlab{a}}), \eprint{hep-ph/0407112}.

\bibitem[{\citenamefont{Ma}(2004)}]{matbmwithvevs}
\bibinfo{author}{\bibfnamefont{E.}~\bibnamefont{Ma}} (\bibinfo{year}{2004}),
  \eprint{hep-ph/0404199}.

\bibitem[{\citenamefont{Lavoura}(2000)}]{LavLmodel2}
\bibinfo{author}{\bibfnamefont{L.}~\bibnamefont{Lavoura}},
  \bibinfo{journal}{Phys. Rev.} \textbf{\bibinfo{volume}{D62}},
  \bibinfo{pages}{093011} (\bibinfo{year}{2000}), \eprint{hep-ph/0005321}.

\bibitem[{\citenamefont{Low and Volkas}(2003)}]{nogo1}
\bibinfo{author}{\bibfnamefont{C.~I.} \bibnamefont{Low}} \bibnamefont{and}
  \bibinfo{author}{\bibfnamefont{R.~R.} \bibnamefont{Volkas}},
  \bibinfo{journal}{Phys. Rev.} \textbf{\bibinfo{volume}{D68}},
  \bibinfo{pages}{033007} (\bibinfo{year}{2003}), \eprint{hep-ph/0305243}.

\bibitem[{\citenamefont{Low}(2004)}]{nogo2}
\bibinfo{author}{\bibfnamefont{C.~I.} \bibnamefont{Low}},
  \bibinfo{journal}{Phys. Rev.} \textbf{\bibinfo{volume}{D70}},
  \bibinfo{pages}{073013} (\bibinfo{year}{2004}), \eprint{hep-ph/0404017}.

\bibitem[{\citenamefont{Weinberg}(1979)}]{dim5}
\bibinfo{author}{\bibfnamefont{S.}~\bibnamefont{Weinberg}},
  \bibinfo{journal}{Phys. Rev. Lett.} \textbf{\bibinfo{volume}{43}},
  \bibinfo{pages}{1566} (\bibinfo{year}{1979}).

\bibitem[{\citenamefont{Cheng and Li}(1980)}]{triplet}
\bibinfo{author}{\bibfnamefont{T.~P.} \bibnamefont{Cheng}} \bibnamefont{and}
  \bibinfo{author}{\bibfnamefont{L.-F.} \bibnamefont{Li}},
  \bibinfo{journal}{Phys. Rev.} \textbf{\bibinfo{volume}{D22}},
  \bibinfo{pages}{2860} (\bibinfo{year}{1980}).

\bibitem[{\citenamefont{Grimus et~al.}(2004{\natexlab{b}})\citenamefont{Grimus,
  Joshipura, Lavoura, and Tanimoto}}]{grimlavtextures}
\bibinfo{author}{\bibfnamefont{W.}~\bibnamefont{Grimus}},
  \bibinfo{author}{\bibfnamefont{A.~S.} \bibnamefont{Joshipura}},
  \bibinfo{author}{\bibfnamefont{L.}~\bibnamefont{Lavoura}}, \bibnamefont{and}
  \bibinfo{author}{\bibfnamefont{M.}~\bibnamefont{Tanimoto}},
  \bibinfo{journal}{Eur. Phys. J.} \textbf{\bibinfo{volume}{C36}},
  \bibinfo{pages}{227} (\bibinfo{year}{2004}{\natexlab{b}}),
  \eprint{hep-ph/0405016}.

\bibitem[{\citenamefont{Zhou and Xing}(2004)}]{xingtext1}
\bibinfo{author}{\bibfnamefont{S.}~\bibnamefont{Zhou}} \bibnamefont{and}
  \bibinfo{author}{\bibfnamefont{Z.-z.} \bibnamefont{Xing}}
  (\bibinfo{year}{2004}), \eprint{hep-ph/0404188}.

\bibitem[{\citenamefont{Xing and Zhang}(2003)}]{xingtext2}
\bibinfo{author}{\bibfnamefont{Z.-z.} \bibnamefont{Xing}} \bibnamefont{and}
  \bibinfo{author}{\bibfnamefont{H.}~\bibnamefont{Zhang}},
  \bibinfo{journal}{Phys. Lett.} \textbf{\bibinfo{volume}{B569}},
  \bibinfo{pages}{30} (\bibinfo{year}{2003}), \eprint{hep-ph/0304234}.

\bibitem[{\citenamefont{Bando and Obara}(2003)}]{bandotext1}
\bibinfo{author}{\bibfnamefont{M.}~\bibnamefont{Bando}} \bibnamefont{and}
  \bibinfo{author}{\bibfnamefont{M.}~\bibnamefont{Obara}},
  \bibinfo{journal}{Prog. Theor. Phys.} \textbf{\bibinfo{volume}{109}},
  \bibinfo{pages}{995} (\bibinfo{year}{2003}), \eprint{hep-ph/0302034}.

\bibitem[{\citenamefont{Bando et~al.}(2004)\citenamefont{Bando, Kaneko, Obara,
  and Tanimoto}}]{bandotext2}
\bibinfo{author}{\bibfnamefont{M.}~\bibnamefont{Bando}},
  \bibinfo{author}{\bibfnamefont{S.}~\bibnamefont{Kaneko}},
  \bibinfo{author}{\bibfnamefont{M.}~\bibnamefont{Obara}}, \bibnamefont{and}
  \bibinfo{author}{\bibfnamefont{M.}~\bibnamefont{Tanimoto}},
  \bibinfo{journal}{Phys. Lett.} \textbf{\bibinfo{volume}{B580}},
  \bibinfo{pages}{229} (\bibinfo{year}{2004}), \eprint{hep-ph/0309310}.

\bibitem[{\citenamefont{Lavoura}(2004)}]{lavtext}
\bibinfo{author}{\bibfnamefont{L.}~\bibnamefont{Lavoura}}
  (\bibinfo{year}{2004}), \eprint{hep-ph/0411232}.

\bibitem[{\citenamefont{Joshipura}(2004)}]{joshtheta13}
\bibinfo{author}{\bibfnamefont{A.~S.} \bibnamefont{Joshipura}}
  (\bibinfo{year}{2004}), \eprint{hep-ph/0411154}.

\bibitem[{\citenamefont{Chen et~al.}(2004)\citenamefont{Chen, Frigerio, and
  Ma}}]{matheta13}
\bibinfo{author}{\bibfnamefont{S.-L.} \bibnamefont{Chen}},
  \bibinfo{author}{\bibfnamefont{M.}~\bibnamefont{Frigerio}}, \bibnamefont{and}
  \bibinfo{author}{\bibfnamefont{E.}~\bibnamefont{Ma}} (\bibinfo{year}{2004}),
  \eprint{hep-ph/0412018}.

\bibitem[{\citenamefont{Petcov and Rodejohann}(2004)}]{rojL}
\bibinfo{author}{\bibfnamefont{S.~T.} \bibnamefont{Petcov}} \bibnamefont{and}
  \bibinfo{author}{\bibfnamefont{W.}~\bibnamefont{Rodejohann}}
  (\bibinfo{year}{2004}), \eprint{hep-ph/0409135}.

\bibitem[{\citenamefont{Grimus and Lavoura}(2004)}]{lavL}
\bibinfo{author}{\bibfnamefont{W.}~\bibnamefont{Grimus}} \bibnamefont{and}
  \bibinfo{author}{\bibfnamefont{L.}~\bibnamefont{Lavoura}}
  (\bibinfo{year}{2004}), \eprint{hep-ph/0410279}.

\bibitem[{\citenamefont{Branco et~al.}(1989)\citenamefont{Branco, Grimus, and
  Lavoura}}]{grimlavL}
\bibinfo{author}{\bibfnamefont{G.~C.} \bibnamefont{Branco}},
  \bibinfo{author}{\bibfnamefont{W.}~\bibnamefont{Grimus}}, \bibnamefont{and}
  \bibinfo{author}{\bibfnamefont{L.}~\bibnamefont{Lavoura}},
  \bibinfo{journal}{Nucl. Phys.} \textbf{\bibinfo{volume}{B312}},
  \bibinfo{pages}{492} (\bibinfo{year}{1989}).

\bibitem[{\citenamefont{Barbieri et~al.}(1998)\citenamefont{Barbieri, Hall,
  Smith, Strumia, and Weiner}}]{barbieriL}
\bibinfo{author}{\bibfnamefont{R.}~\bibnamefont{Barbieri}},
  \bibinfo{author}{\bibfnamefont{L.~J.} \bibnamefont{Hall}},
  \bibinfo{author}{\bibfnamefont{D.~R.} \bibnamefont{Smith}},
  \bibinfo{author}{\bibfnamefont{A.}~\bibnamefont{Strumia}}, \bibnamefont{and}
  \bibinfo{author}{\bibfnamefont{N.}~\bibnamefont{Weiner}},
  \bibinfo{journal}{JHEP} \textbf{\bibinfo{volume}{12}}, \bibinfo{pages}{017}
  (\bibinfo{year}{1998}), \eprint{hep-ph/9807235}.

\bibitem[{\citenamefont{Petcov}(1982)}]{petcovL}
\bibinfo{author}{\bibfnamefont{S.~T.} \bibnamefont{Petcov}},
  \bibinfo{journal}{Phys. Lett.} \textbf{\bibinfo{volume}{B110}},
  \bibinfo{pages}{245} (\bibinfo{year}{1982}).

\bibitem[{\citenamefont{Leung and Petcov}(1983)}]{leungL}
\bibinfo{author}{\bibfnamefont{C.~N.} \bibnamefont{Leung}} \bibnamefont{and}
  \bibinfo{author}{\bibfnamefont{S.~T.} \bibnamefont{Petcov}},
  \bibinfo{journal}{Phys. Lett.} \textbf{\bibinfo{volume}{B125}},
  \bibinfo{pages}{461} (\bibinfo{year}{1983}).

\bibitem[{\citenamefont{Joshipura and Rindani}(2000)}]{joshipuraL}
\bibinfo{author}{\bibfnamefont{A.~S.} \bibnamefont{Joshipura}}
  \bibnamefont{and} \bibinfo{author}{\bibfnamefont{S.~D.}
  \bibnamefont{Rindani}}, \bibinfo{journal}{Eur. Phys. J.}
  \textbf{\bibinfo{volume}{C14}}, \bibinfo{pages}{85} (\bibinfo{year}{2000}),
  \eprint{hep-ph/9811252}.

\bibitem[{\citenamefont{Fogli et~al.}(2004)}]{fog04}
\bibinfo{author}{\bibfnamefont{G.~L.} \bibnamefont{Fogli}} \bibnamefont{et~al.}
  (\bibinfo{year}{2004}), \eprint{hep-ph/0408045}.

\bibitem[{\citenamefont{Frigerio and Smirnov}(2003)}]{rgesmirnov}
\bibinfo{author}{\bibfnamefont{M.}~\bibnamefont{Frigerio}} \bibnamefont{and}
  \bibinfo{author}{\bibfnamefont{A.~Y.} \bibnamefont{Smirnov}},
  \bibinfo{journal}{JHEP} \textbf{\bibinfo{volume}{02}}, \bibinfo{pages}{004}
  (\bibinfo{year}{2003}), \eprint{hep-ph/0212263}.

\bibitem[{\citenamefont{Antusch et~al.}(2001)\citenamefont{Antusch, Drees,
  Kersten, Lindner, and Ratz}}]{rgemunich}
\bibinfo{author}{\bibfnamefont{S.}~\bibnamefont{Antusch}},
  \bibinfo{author}{\bibfnamefont{M.}~\bibnamefont{Drees}},
  \bibinfo{author}{\bibfnamefont{J.}~\bibnamefont{Kersten}},
  \bibinfo{author}{\bibfnamefont{M.}~\bibnamefont{Lindner}}, \bibnamefont{and}
  \bibinfo{author}{\bibfnamefont{M.}~\bibnamefont{Ratz}},
  \bibinfo{journal}{Phys. Lett.} \textbf{\bibinfo{volume}{B519}},
  \bibinfo{pages}{238} (\bibinfo{year}{2001}), \eprint{hep-ph/0108005}.

\end{thebibliography}

\end{document}